\documentclass[%
reprint,
superscriptaddress,
amsmath,amssymb,
aps,
prl,
floatfix,
]{revtex4-1}

\usepackage{graphicx}
\usepackage{dcolumn}
\usepackage{bm}
\usepackage{braket}
\usepackage[german,english]{babel}
\usepackage[colorlinks=true,linkcolor=blue,citecolor=blue,breaklinks]{hyperref}
\usepackage[pdftex]{color}
\usepackage{ulem}
\usepackage{comment}
\usepackage{natbib}
\usepackage{xcolor}
\begin{document}
	
\title{Observing Quasiparticles through the Entanglement Lens}
\author{Yizhi You}
\affiliation{Princeton Center for Theoretical Science, Princeton University, NJ, 08544, USA}
\author{Elisabeth Wybo}
\affiliation{Department of Physics, Technical University of Munich, 85748 Garching, Germany}
\affiliation{Munich Center for Quantum Science and Technology (MQCST), D-80799 Munich, Germany}
\author{Frank Pollmann}
\affiliation{Department of Physics, Technical University of Munich, 85748 Garching, Germany}
\affiliation{Munich Center for Quantum Science and Technology (MQCST), D-80799 Munich, Germany}
\author{S. L. Sondhi}
\affiliation{Department of Physics, Princeton University, NJ 08544, USA}
\date{\today}
	
\begin{abstract}
The low energy physics of interacting quantum systems is typically understood through the identification of the relevant quasiparticles or low energy excitations and their quantum numbers. We present a quantum information framework that goes beyond this to examine the nature of the entanglement in the corresponding quantum states. We argue that the salient features of the quasiparticles, including their quantum numbers, locality and fractionalization are reflected in the entanglement spectrum and in the mutual information. We illustrate these ideas in the specific context of the $d=1$ transverse field Ising model with an integrability breaking perturbation.

\end{abstract}

\maketitle
	
\textbf{Introduction---} Our understanding of quantum many body theory has been deeply affected by the quantum information ``revolution'' which has, among other ideas, led to the systematic study of the entanglement present in quantum states \cite{RevModPhys.80.517}. The most prominent product of this study is the 
formulation of the area law for the entanglement entropy (EE), i.e., the von Neumann entropy $S=-\mathrm{Tr}\rho_A\log\rho_A$ of the reduced density matrix $\rho_A$ of a subsystem $A$, in ground states of local quantum systems \cite{Hastings:2007,Srednick1993}. That in turn is the basis for an entire family of tensor network methods for representing and computing with ground states which has revolutionized the computational study of quantum many body systems \cite{Fannes-1992,Verstraete9,Schollwoeck11}. Moreover, the area law for the EE has led to the identification of a universal subleading correction, the topological EE, which is a manifestation of topological order \cite{KitaevPreskill,Levin-2006}. Moving beyond the entropy, the entanglement spectrum (ES) has been widely explored  \cite{Li-2008,Peschel_2009,pollmann2012symmetry,PhysRevB.84.195103,PhysRevLett.105.115501,PhysRevLett.104.130502,chandran2014universal,PhysRevLett.110.236801}. Altogether the study of entanglement has provided a new route to exploring quantum phases and phase transitions, and has been very successful in describing a wide variety of exotic states and related phase transitions \cite{calabrese2008entanglement,ding2009entanglement,zhang2012quasiparticle}.

In this paper we wish to use the entanglement lens to look at very low energy dynamics. In this limit, as is well known, the dynamics in phases can be typically be understood in terms of the elementary excitations of the system. Specifically we wish to look at quasiparticles (QP) which we will take to be gapped particle-like excitations as opposed to collective excitations like Goldstone bosons\cite{metlitski2011entanglement}.
There is however no expectation, as in traditional many-body physics, that quasiparticles necessarily have an adiabatic continuation to particle-like excitations of a free system. In particular, we also consider QPs that are fractionalized and/or cannot be created by local operators (e.g., spinons or solitons). In doing so, we also move beyond a body of results on excitations in integrable models \cite{castro2018entanglement,eisler2019front,castro2019entanglement,jafarizadeh2019bipartite,castro2018entanglementa} which prefigures some of our work although even where our results overlaps theirs, we are interested in their properties in generic, non-integrable, systems.

In what follows we study the entanglement structure of quasiparticle states. The questions of interest are: How is the entanglement present in the ground state modified by the presence of the quasiparticle? Are the quantum numbers of the quasiparticle reflected in the entanglement spectrum of the quasiparticle states? Can we distinguish quasiparticles created by local operators from those created by non-local operators via the quantum information content of the quasiparticle states? To this end we will present results on the $d=1$ transverse field Ising model with an integrability breaking term added. In the symmetric/paramagnetic phase the quasiparticle is local while in the broken symmetry/ferromagnetic phase the quasiparticle is a domain wall/soliton and thus is created by a non-local operator. In the symmetric phase we find the EE exhibits a term above the ground state contribution that arises from the uncertainty in the position of the quasiparticle. More finely, we find an exact doublet structure in the entanglement spectrum for a symmetric bipartition that we trace to a combination of the symmetry of the partition and the existence of a $Z_2$ charge carried by the quasiparticle. In the broken symmetry phase the EE of the domain wall excitation exhibits an additional universal contribution that reflects the underlying broken symmetry. As the $Z_2$ charge is no longer well defined, the ES no longer has an exact spectral degeneracy. Finally, we show that the non-locality of the domain wall excitation can be detected by computing the long-range mutual information(MI) between spatially separated qu-bits in the corresponding quantum state.
We briefly discuss generalizations of these ideas to symmetry fractionalization and topological QP in higher dimensions which will be presented in future publications.

\textbf{Entanglement in a local QP --}
Our exemplar of a local QP will arise in the paramagnetic/symmetric phase of the $d=1$ transverse field Ising model:
  \begin{align} 
  &H= \sum_i J_z \sigma^z_i\sigma^z_{i+1}+h_x \sigma^x_i+J_x \sigma^x_i\sigma^x_{i+1}
  \end{align}
where we have added a weak integrability breaking interaction $J_x\sigma^x_i\sigma^x_{i+1}$ for
genericity. When $h_x> J_z + O(J_x)$, the system is in the paramagnetic phase. In the extreme limit $h_x \gg J_z$ the ground state is fully polarized in the $x$ direction and the QP---which we will refer to as a $Z_2$ magnon---is generated by local spin flip operator $Q_{i}=\sigma^z_i$ which creates a $Z_2$ charge measured by the parity operator $P=\prod_i \sigma^x_i$. Away from the extreme paramagnetic limit $Q_i$ remains a $Z_2$ odd operator but now spreads out over a correlation length $\xi$. 
With periodic boundary conditions, the system is translationally invariant and the low energy magnon states are momentum eigenstates of the form \cite{castro2018entanglement}, 
\begin{align} 
  &|\psi^{1M}\rangle_k = \sum_i e^{i k r_i}Q_i |\mathrm{GS} \rangle
  \end{align}
where we have ignored the subleading $k$ dependence of the operator $Q_i$ itself. With open boundary conditions we will have standing wave analogs thereof.

\begin{figure}[h!]
	\includegraphics[width=1.0\columnwidth]{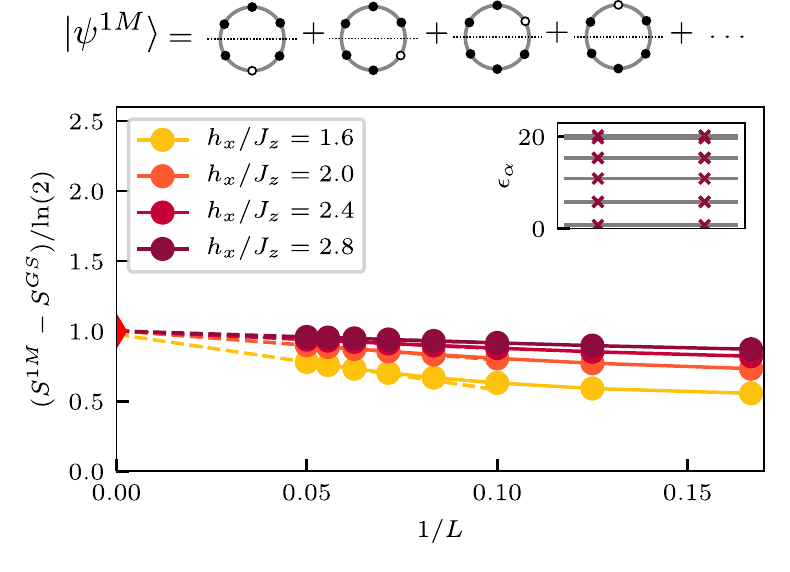}
	\caption{Difference of the EE between the ground state and the single magnon excitation in the paramagnetic phase of the transverse field Ising model with $J_x = 0.1$ for a reflection symmetric bipartition. The red marker on the y-axis shows the expected $\ln(2)$ contribution. Inset: The entanglement spectrum exhibits an inversion symmetry protected two-fold degeneracy.}
	\label{fig:pm}
\end{figure}

We will now consider a spatial bipartition into regions of length $pL$ and $(1-p)L$.
\begin{align}
S^{1M}_{p}=S^{\mathrm{GS}}_{p}+p\ln(p)+(1-p)\ln(1-p)+O(1/L) \ ,
\label{qp}
\end{align}
irrespective of $k$. The additional EE due to the presence of the magnon over the ground state value originates from the uncertainty in its location relative to the cuts. For a finite correlation length, there is a finite probability of the magnon reaching across the cut, yielding $O(1/L) $ corrections. For a symmetric cut in the thermodynamic limit, the additional EE created by QP is $\ln(2)$---one bit---as illustrated in Fig.~\ref{fig:pm}.

We now consider the magnon entanglement spectrum, defined as $\epsilon_{\gamma} = -\log \lambda^2_{\gamma}$, which is obtained from a Schmidt decomposition 
\begin{equation}
|\psi^{1M}\rangle= \sum_{\gamma} \lambda_{\gamma}|\gamma\rangle_A |\gamma\rangle_B.
\end{equation}
Note that $\lambda_{\gamma}^2$ are the eigenvalues of the reduced density matrices $\rho_A$ and $\rho_B$ for the two subsystems. 
For a reflection symmetric cut (see inset in Fig.~\ref{fig:pm}), the entanglement spectrum exhibits an {\it exact} two-fold degeneracy.
We now show that this exact degeneracy of the Schmidt values arises from a combination of the reflection symmetry $R|\psi^{1M} \rangle=|\psi^{1M} \rangle$ and the  non-trivial $Z_2$ charge of the magnon $P|\psi^{1M} \rangle=-|\psi^{1M} \rangle$:
To this let us assume instead that the entanglement spectrum contains a non-degenerate eigenvalue $\lambda_{\gamma}$.
The state $|\psi^{1M}\rangle$ is symmetric under $R$ and the Schmidt decomposition is unique up to unitary rotations among degenerate eigenspaces. Thus it must hold that  $|\gamma\rangle_A =R|\gamma\rangle_B$ upto a U(1) phase.
Moreover, for a parity symmetric  state, the Schmidt states are eigenstates with $P|\gamma\rangle_{A/B}=\pm |\gamma\rangle_{A/B}$.
Since $P|\psi^{1M} \rangle=-|\psi^{1M} \rangle$,  $|\gamma\rangle_A$ and $|\gamma\rangle_B$ must have opposite parity eigenvalues.%
This leads to a contradiction with the requirement $|\gamma\rangle_A =R|\gamma\rangle_B$ as $R$ does not change the charge.
Consequently, all Schmidt values have to be degenerate \footnote{The arguments here can be generalized to single magnon states with $k \ne 0$ where instead of inversion we can use translations by $L/2$. }.
\\

Our discussion above can be readily generalized to $Z_M$ paramagnetic phases. For symmetric bipartitions, the quasiparticle carrying unit charge (modulo $M$) in the $Z_M$ transverse Ising model again contributes an additional EE $\ln(2)$ in the large system limit due to its position uncertainty. However,  the entanglement spectrum exhibits an exact two-fold degeneracy for finite system sizes only if $M$ is even as we discuss in the Supplementary Material.

\begin{figure}[h!]
	\includegraphics[width=1.0\columnwidth]{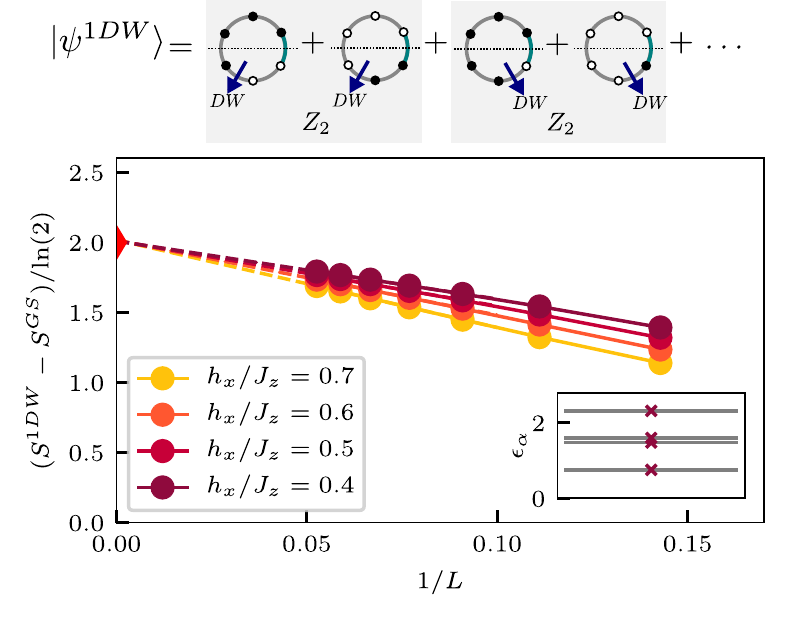}
	\caption{Difference of the EE between the symmetry broken ground state and the single domain wall excitation in the ferromagnetic phase of the transverse field Ising model with $J_x = 0.1$. We use anti-periodic boundary condition and choose a reflection symmetric bipartition. The red marker on the y-axis shows the expected $2\ln(2)$ contribution. Inset: The entanglement spectrum does not exhibit any symmetry protected degeneracies.}
	\label{fig:fm}
\end{figure}

\textbf{Entanglement in non-local QPs--} We now move to the FM phase of the transverse field Ising model whose $Z_2$ symmetry breaking ground states are two-fold degenerate in the infinite system limit. The elementary excitations now are domain walls which are topological defects between $|\uparrow \rangle_z$ and $|\downarrow\rangle_z$ patterns. Creating a domain wall is clearly a non-local operation involving
flipping all the spins on the left(or right) side of of the intended domain wall location. As 
we would like to study finite systems, boundary conditions now become important. With periodic boundary conditions, the ground states take the form of macroscopic superpositions of the $|\uparrow \rangle_z$ and $|\downarrow\rangle_z$ states (``cat states'') which are eigenstates with $P=\pm 1$. Separated from these via a gap we will find states with pairs of domain walls. As this is a somewhat complicated situation to think about, let us begin the simpler case of the FM Ising model with anti-periodic boundary conditions whose ground state contains a single domain wall.
As the eigenstate preserves the $Z_2$ symmetry, for each configuration with fixed domain wall locations, there are two globally distinct patterns related by the $Z_2$ symmetry as illustrated in Fig.~\ref{fig:fm}. Consequently, its EE with respect to the partial chain with length $pL$ takes the form,
\begin{align}
S^{\mathrm{1DW}}_{p}=S^{\mathrm{GS}}_{p}+p\ln(p)+(1-p)\ln(1-p)+\ln(2)+O(1/L)
\label{tw}
\end{align}
where we assume that $pL$ is much larger than the correlation length. The EE of a single domain wall excitation has three contributions: $S^{\mathrm{GS}}_{p}$ is the local EE of the symmetry broken Ising FM ground state with PBC which is non-universal and dependent on the local entanglement pattern near the cut. The $p$ dependent
contribution arises from the domain wall location uncertainty.
The final $\ln(2)$ piece is the universal EE due to the macroscopic superposition for each fixed domain wall location. Provided the domain wall excitation is invariant under $Z_2$, the cat-state with globally distinct patterns persists so the $\ln(2)$ contribution is robust against any perturbation. For a spatial symmetric bipartition, the additional EE generated by domain wall is then $2 \ln(2)$ as one sees in the data in Fig.~\ref{fig:fm}.
We note that the entanglement spectrum of the symmetric bipartition does {\it not} exhibit an exact two-fold degeneracy. This lack of degeneracy arise from the fact that the domain wall does not carry a well defined $Z_2$ quantum number in the phase where the symmetry is broken.
 
Returning to the case of PBCs we might expect that the two domain wall states will exhibit 
an additional EE
\begin{align}
&S^{2\mathrm{DW}}_{p}-S^{\mathrm{GS}}_{p}=f^2(p)+\ln(2) 
\label{2DW}
\end{align}
where $f^2(p)$ is the contribution from the joint location uncertainty of the two excitations.
In the zero-correlation length limit, we expect $f^2(p)=p^2\ln(p^2)+(1-p)^2\ln((1-p)^2)$. The second term in Eq.~(\ref{2DW}) is the universal $\ln(2)$ EE discussed above.
Such a result would again indicate that non-local quasiparticles in symmetry breaking phases carry two types of information entropy, one from their positional uncertainty and the other universal piece from their non-locality. Unfortunately, away from this limit the explicit form of $f^2(p)$ is non-universal and varies between different eigenstates while depending on microscopic details such as the QP interaction and correlation length. To extract the universal $\ln(2)$ pieces, we consider the reduced density matrix of a small region with $p\ll1$ so the positional uncertainty part almost vanishes and the EE converges to the universal $\ln(2)$. The spirit of the form in Eq.~(\ref{2DW}) reveals that the additional EE $S^{\mathrm{DW}}_{p}-S^{\mathrm{GS}}_{p}$ carried by the QP states is composed of two additive contributions from the positional uncertainty and the macroscopic superposition respectively. The second, universal, piece will generalize \footnote{We only consider discrete symmetry breaking with a gapped spectrum and finite ground state manifold} and depends on the dimension of the ground state manifold in the symmetry breaking phase, e.g. for the $Z_M$ FM Ising model it becomes $\ln(M)$.

\begin{figure}[h!]
	\includegraphics[width=1.0\columnwidth]{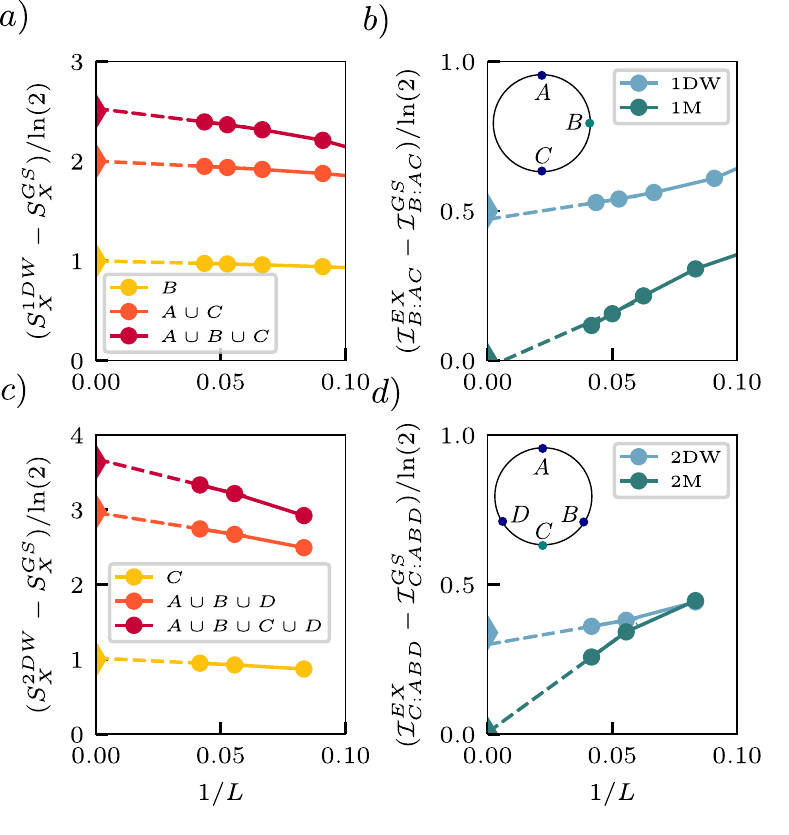}
	\caption{Mutual information for differnt partitions of the chain, differentiating the cases of magnon and domain-wall excitations, in the para- ($h_x/J_z = 1.6$) and ferromagnetic ($h_x/J_z = 0.2$) phase, respectively. Panels a) and b) show the tripartite mutual information and its contributions for single quasi-particle excitations. Panels c) and d) show the quadripartite mutual information and its contributions for exicitations of two quasiparticles. The markers on the y-axes show the expected contribution in the thermodynamic limit.}
	\label{fig:mi}
\end{figure}

\textbf{LRMI in non-local QPs--} Following the identification of this universal entanglement contribution, we propose another intimately related entanglement property, the long-range mutual information (LRMI) \cite{jian2015long} between spatially widely separated qubits stemming from the presence of one or more non-local QPs. We again begin with the state containing single domain wall induced by anti-periodic boundary conditions. As noted above, the EE of a single spin $B$ is,
\begin{align}
S^{1\mathrm{DW}}_{B}=\ln(2) \ .
\end{align}
Next we consider the EE of two isolated spins $A,C$ separated by $L/2$ as in Fig.~\ref{fig:mi}(a) and (b), and find
\begin{align}
S^{1\mathrm{DW}}_{A\cup C}=2\ln(2)-O(1/L) \ .
\end{align}
Due to the uncertainty in the location of the domain wall, the two spins can be parallel or anti-parallel which produces $\ln(2)$ EE in addition to the universal EE of $\ln(2)$ from the global superposition in a cat state.

Now we place $B$ in the middle of $A,C$ and calculate the EE for the three isolated spins,
\begin{align}
S^{1\mathrm{DW}}_{A\cup C\cup B}=5/2\ln(2)-O(1/L)
\end{align}

Finally, the mutual information between $B$ and $A\cup C$ is
\begin{align}
I_{B:AC}=-S^{1\mathrm{DW}}_{A\cup C\cup B}+S^{1\mathrm{DW}}_{A\cup C}+S^{1\mathrm{DW}}_{B}=1/2\ln(2)
\label{mi2}
\end{align}
Such a non-vanishing MI implies that a measurement of $A\cup C$ would reduce the information carried by $B$. This is obvious in the sense that once $A, C$ are parallel to each other, the domain wall is outside so the spin $B$ configuration is fixed to be parallel to $A, C$. In general, the mutual information between the set of three spins $B$ and $A,C$ is always non-vanishing provided each pair of them is far apart.

For QP states containing a pair of domain wall excitation, we still find a non-vanishing mutual information to appear between three isolated spins and the spin in the middle shown in Fig.~\ref{fig:mi}(c) and (d).

\begin{align}
I_{C:ABD}=S^{2\mathrm{DW}}_{A\cup B\cup D}+S^{2\mathrm{DW}}_{C}-S^{2\mathrm{DW}}_{A\cup B\cup C\cup D}=\ln(2)-g
\label{eq:mi4}
\end{align}
Here $g$ is a non-universal term which depends on the spatial distribution of the two domain walls and their interaction with an upper limit $g<\frac{7}{9}\ln(2)$ (See Supplementary Material for details). 
Subsequently, the mutual information is bounded from below.
This non-vanishing LRMI can be understood as follows: Due to the non-locality of the DW excitations, the entropy of a single spin can be strongly reduced when performing a measurement or projection to the spins infinite far away, $e.g.$, if we project $(A,B,D)$ to the $(1,0,0)$ pattern, the region between $B,D$ are nearly polarized so $C$ is pinned into the $0$ state.

The LRMI between distant regions generates an experimentally accessible protocol to probe different types of QPs via partial measurement of conditional EE \cite{ben2020disentangling}. Namely, a measurement of subsystem A leads to a dramatic reduction of the bipartite EE of region B even though they are spatially separated apart. In particular, the non-vanishing mutual information we explored in Eqs.~(\ref{mi2}) and~(\ref{eq:mi4}) can be extended to Renyi-entropies which are directly experimentally accessible \cite{PhysRevLett.109.020504,Kaufman794,PhysRevLett.109.020505,PhysRevLett.120.050406,choo2018measurement}. 

\textbf{Generalizations and Outlook--}
While we introduced our framework by considering eigenstates of the transverse field Ising chain as concrete examples, the entanglement lens can be applied to characterize QPs in a much broader sense\cite{YY}: 
For example, QPs can carry fractionalized quantum numbers, as it is for example the case in the Majumdara-€"Ghosh model \cite{ghosh}, where domain walls between different dimerization patterns are $S=1/2$ spinons.
In this case, the symmetry fractionalization leads to additional characteristic pattern in the entanglement structure of the QP excitations.
Moreover, we expect that QP excitations in symmetry protected topological (SPT) phases \cite{Pollmann2010,Chen2011-et,you2018subsystem} still exhibit conditional mutual information despite the fact that the characteristic string order \cite{denNijsRommelse} is only present in the ground state and disappears in QP states.
Going beyond the one-dimensional case, it is an exciting question how to characterize anyonic QPs that emerge in topologically ordered systems \cite{wen1990topological}.
Here the non-local nature of the QP is expected to lead to long range mutual information that allow to characterize the underlying entanglement structure.
Another exciting direction is to investigate the entanglement dynamics of individual QP excitations.
Imagine we begin with an initial state by adding a static QP at site-$i$ to the ground state.
The QP will delocalize due to the unitary time evolution and the entanglement growth after the quench turns out to differ qualitatively between local and non-local QP. 

\textbf{Conclusions--}
We proposed a unified way to characterize QPs using the EE and mutual information.
We identified universal properties that provide a powerful framework to characterize one-dimensional symmetric and symmetry broken phases of matter via their local and non-local QP excitations, respectively. 
Thus the low energy QP state  inherits the underlying symmetry structure of the ground state. 
Broadly speaking, our approach provides a new route to probe QP and collective excitations in complex quantum materials.

{\em Acknowledgments.---}We would like to thank Ruben Verresen for stimulating discussions. FP is funded by the European Research Council (ERC) under the European Unions Horizon 2020 research and innovation program (grant agreement No. 771537). FP acknowledges the support of the DFG Research Unit FOR 1807 through grants no. PO 1370/2-1, TRR80, and the Deutsche Forschungsgemeinschaft (DFG, German Research Foundation) under Germany's Excellence Strategy EXC-2111-390814868. YY and FP acknowledge the support from Banff International Research Station where this work was partly initiated. SLS acknowledges support from the United StatesDepartment  of  Energy  via  grant  No.   DE-SC0016244.

\clearpage

\setcounter{figure}{0}
\setcounter{section}{0}
\setcounter{equation}{0}
\renewcommand{\theequation}{S\arabic{equation}}
\renewcommand{\thefigure}{S\arabic{figure}}

\onecolumngrid

\newcommand{\vsigma}{\mbox{\boldmath $\sigma$}}

\section*{\Large{Supplemental Material}}

\section{Universal EE for non-local QP}
We elucidate the origin of universal $\ln(2)$ EE for domain wall excitations with respect to different spatial partitions. For FM Ising model with anti-periodic boundary conditions and a single domain wall fluctuating in the bulk, its entanglement eigenvectors with respect to the partial chain with length $pL$ are,
\begin{align}
& q_1=|0...1 \rangle, q_2=|1...0 \rangle,
q_3=|0...0 \rangle, q_4=|1...1 \rangle,
\end{align}
Here $q_1,q_2$ label the states carrying a fluctuating domain wall inside the region. $q_1,q_2$ label the states carrying a fluctuating domain wall outside the region. As the reduced density matrix is invariant under $Z_2$ symmetry, we require $q_1,q_2$ patterns to be degenerate in the entanglement spectrum and same applies to $q_3,q_4$. This indicates there exist two distinct patterns with the same domain wall location in the reduced density matrix. Subsequently, we obtain additional universal $\ln(2)$ contribution to the EE.

\section{Mutual information lower bound}

For the QP state with a pair of domain wall excitations, the non-vanishing mutual information appears between three isolated spins and the spin in the middle as illustrated in the inset of Fig.~\ref{fig:mi}(d).

\begin{align}
I_{C:ABD}=S^{\mathrm{DW}}_{A\cup B\cup D}+S^{\mathrm{DW}}_{C}-S^{\mathrm{DW}}_{A\cup B\cup C\cup D}=\ln(2)-g
\end{align}

Here $g$ is a non-universal piece which indicates the additional entropy generated by the location uncertainty of the domain wall if we include the additional $C$ spin between $B,D$. Its value depends on the spatial distribution of the two domain walls and their interaction. To demonstrate the non-vanishing mutual information, we need to derive the upper limit of $g$. Generally speaking, adding a spin-1/2 degree of freedom to the reduced density matrix creates additional entropy with a maximum of $\ln(2)$. However,
if there is no domain wall between $B,D$, including the additional $C$ does not create new uncertainty so the entropy does not change. If there is a single domain wall between $A,D$ and another between $A, B$ (with probability $2/9$), the configuration of C is pinned to be parallel to $B,D$ so there is no additional entropy generated.
Based on this estimate of the upper limit, we reach the conclusion that $g<(1-2/9)\ln(2)=7/9\ln(2)$.

\section{$Z_3$ PM phase}

In this section, we generalize our search of QP entanglement to $Z_M$ Ising models. We begin by
examining the QP entanglement spectrum in the in the paramagnetic phase of the $Z_3$ clock model . 
  
  \begin{align} 
  &H= \sum_i \sigma^{\dagger}_i\sigma_{i+1}+h \tau_i,\nonumber\\
  &\tau_i \sigma_j=e^{i2\pi/3}\delta_{ij}\sigma_j \tau_i
  \end{align}

\begin{figure}[h]
\includegraphics[width=0.2\textwidth]{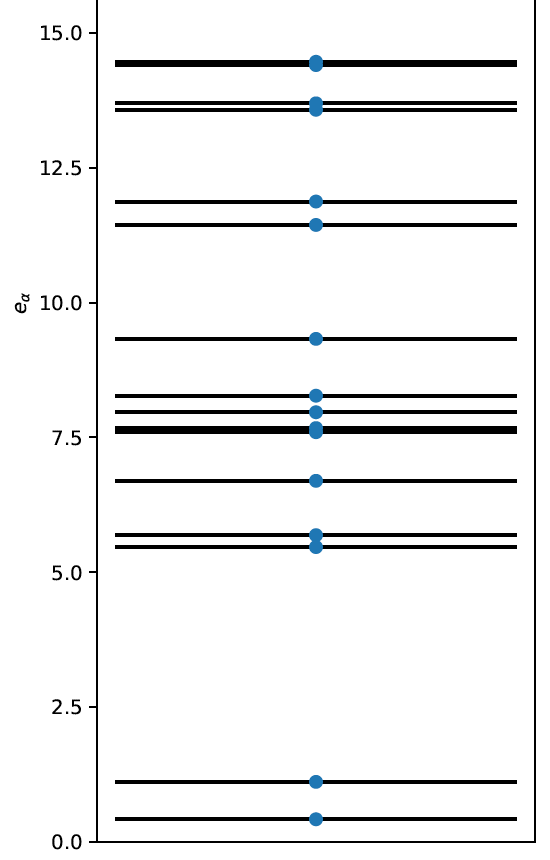}
\caption{ The entanglement spectrum of the zero-momentum QP excited state in $Z_3$ clock model with 12 sites and $h=3.8$.} 
\label{z3case}
\end{figure}

 For the QP state with zero momentum, its wave function carrying charge 1 (modulo 3) is both reflection and $Z_3$ symmetric. However, neither of these symmetries, individually or in combination, renders a projective symmetry for the reduced density matrix with respect to the partial spin chain. In particular, for a center symmetric cut, the QP wave function has the Schmidt decomposition,
\begin{equation}
|\psi^{1M}\rangle= \sum_{\gamma} \lambda_{\gamma}|\gamma\rangle_A |\gamma\rangle_B.
\end{equation}
 One can have an equal configurations as $|\gamma\rangle_A =|\gamma\rangle_B$, each of which carries charge 2(modulo 3). Such configuration still denotes a charge 1 QP state with a unique energy level in the ES.
Hence, the QP ES in $Z_3$ paramagnetic chain does not render any nontrivial degeneracy as we shown in Fig.~\ref{z3case}. The same conclusion applies for other QP states in $Z_{2m+1}$ clock model.

 Likewise, for $Z_{2m}$ Ising PM phase, the QP wavefunction carries odd charge so the reduced density matrix for each half-chain renders a symmetric partition of even and odd charge patterns. This again enforces a two-level degeneracy in the entanglement spectrum.

Despite the lack of an exact degeneracy in the QP entanglement spectrum for the $Z_3$ case, we still expect that the additional EE created by QP converges to $\ln(2)$ due to QP location uncertainty. 

Assume the ground state with net charge has the following Schmidt decomposition,
  \begin{align} 
 &|\mathrm{GS} \rangle= \sum_{\gamma} \lambda'_{\gamma}|\gamma\rangle_A |\gamma\rangle_B. 
 \end{align}
 As the PM state is short-range entangled with a finite correlation length, the Schmidt states $|L\rangle_a$ differ near the cut but share the same pattern far from the cut. Thus, for any local operator $S_i$ in the left half of the chain far from the cut,
  \begin{align} 
 &\mathrm{Tr} [ S_i \rho^L] =\langle \mathrm{GS} |S_i | \mathrm{GS} \rangle
 \end{align}
 
 The QP state can be represented as,
\begin{align} 
  &|\psi^{1M}\rangle_k = \sum_i e^{i k r_i}Q_i |\mathrm{GS} \rangle\nonumber\\
 &= \sum_{\gamma} \lambda'_{\gamma}~[ (\sum_{i \in L} e^{i k r_i}Q_i)|\gamma\rangle_A |\gamma\rangle_B  + |\gamma\rangle_A  \sum_{i \in R} (e^{i k r_i}Q_i )|\gamma\rangle_B]
 \end{align}
 With $Q_i$ being the operator that creates a charged excitation around site-i.
 
 For $Q_i$ far from the cut, 
  \begin{align} 
 &\langle \gamma |_A Q_i|  \gamma'\rangle_A =\langle \mathrm{GS} |Q_i | \mathrm{GS} \rangle=0
 \end{align} 
 Provided $Q_i$ is the QP operator with a finite gap, its ground state expectation value vanishes. Thus, we reach the conclusion that $|\gamma \rangle_A$ and $(\sum_{i \in L} e^{i k r_i}Q_i)|\gamma \rangle_A$ are almost an orthogonal basis with an overlap decaying to zero in the thermodynamic limit (with $O(1/L)$ power-law correction).
This further implies the EE has an additional $\ln(2)$ corresponding to the pattern with or without a quasiparticle in the half-chain.

\vspace{0.7cm}

\end{document}